\documentclass[twocolumn,prl,color,superscriptaddress,psfig,showpacs]{revtex4}

\usepackage{graphicx}
\usepackage{epsfig}
\usepackage{color}
\usepackage{amssymb}
\usepackage{wasysym}

\newcommand{\beq}{\begin{equation}}
\newcommand{\eeq}{\end{equation}}
\newcommand{\Zr}{Li$_2$ZrCuO$_4$}
\newcommand{\li}{Li$_{\rm II}$}
\newcommand{\lii}{Li$_{\rm I}$}

\newcommand{\vl}[1]{\textcolor{black}{#1}}

\begin{document}

\title{Quantum electric dipole glass and frustrated magnetism near a critical
point in \Zr\ \footnote{with some amendments to appear in Europhysics Letters (EPL) {\bf 88} (2009) 27001; http://epljournal.edpsciences.org/} }
\author{E. Vavilova}
\affiliation{Institute for Solid State Research, IFW Dresden, 01171 Dresden, Germany} \affiliation{Zavoisky Physical-Technical Institute of the RAS - 420029, Kazan, Russia}
\author{A.S. Moskvin}
\affiliation{Institute for Theoretical Solid State Physics, IFW Dresden, 01171 Dresden, Germany} \affiliation{Ural State University, 620083, Ekaterinburg, Russia}
\author{Y.C. Arango}
\affiliation{Institute for Solid State Research, IFW Dresden, 01171 Dresden, Germany}
\author{A. Sotnikov}
\affiliation{Institute for Solid State Research, IFW Dresden, 01171 Dresden, Germany} \affiliation{A.\,F. Ioffe Physical-Technical Institute of the RAS, 194021 St.-Petersburg, Russia}
\author{S.-L. Drechsler}
\affiliation{Institute for Theoretical Solid State Physics, IFW Dresden, 01171 Dresden, Germany}
\author{R. Klingeler}
\affiliation{Institute for Solid State Research, IFW Dresden, 01171 Dresden, Germany}
\author{O. Volkova}
\affiliation{Moscow State University, 119992 Moscow, Russia}
\author{A. Vasiliev}
\affiliation{Moscow State University, 119992 Moscow, Russia}
\author{V. Kataev}
\email{v.kataev@ifw-dresden.de}
\affiliation{Institute for Solid State Research, IFW Dresden, 01171 Dresden, Germany}
\author{B. B\"{u}chner}
\affiliation{Institute for Solid State Research, IFW Dresden, 01171 Dresden, Germany}

\date{October 22, 2009}



\begin{abstract}
We report a new peculiar effect of the interaction between a sublattice of frustrated quantum spin-1/2 chains and a sublattice of pseudospin-1/2
centers (quantum electric dipoles) uniquely co-existing in the complex oxide $\gamma$-\Zr\,($\equiv \rm Li_2CuZrO_4$). $^7$Li nuclear magnetic-,
Cu$^{2+}$ electron spin resonance  and a complex dielectric constant data reveal that the sublattice of Li$^+$-derived electric dipoles orders glass
like at \vl{$T_{\mathrm g} \simeq 70$\,K} yielding a spin site nonequivalency  in the CuO$_2$ chains. We suggest that such a remarkable interplay
between electrical and spin degrees of freedom might strongly influence the properties of the spiral spin state in \Zr\ that is close to a quantum
ferromagnetic critical point. In particular that strong quantum fluctuations and/or the glassy behavior of electric dipoles might renormalize the
exchange integrals affecting this way the pitch angle of the spiral as well as be responsible for the missing multiferroicity present in other
helicoidal magnets.
\end{abstract}

\maketitle

The recently discovered quantum spin chain compound \Zr\
\cite{Drechsler1} is a strongly frustrated quantum magnet located
on the spin spiral side of the magnetic phase diagram in the very
vicinity of a quantum critical point (QCP) to ferromagnetism.
Surprisingly, contrarily to other related materials where
ferroelectricity occurs simultaneously with a spiral magnetic
order yielding a multiferroic behavior
\cite{Fiebig,Naito07,Park07}, here, below the magnetic ordering at
$T_N \approx 6$\,K {\it no} multiferroicity has been found
hitherto \cite{Tarui08}.  This is puzzling since the usually
supporting features such as the low symmetry of buckled CuO$_2$
chains allowing Dzyaloshinski-Moriya interactions and/or special
positions for impurity spins are present in \Zr. In addition, the
unusually large magnitude of the estimated main (nearest neighbor)
exchange integral $J_1 \sim -300$\,K \cite{Drechsler1} differs
markedly from those of other related edge-shared CuO$_2$ chain
compounds such as Li$_2$CuO$_2$ ($J_1 \approx -230$\,K)
\cite{Lorenz} and Ca$_2$Y$_2$Cu$_5$O$_{10}$ as well as from
L(S)DA+$U$ calculations for \Zr itself \cite{Schmitt09}. The
microscopic reason for these puzzling deviations are completely
unclear at present. We conjecture that under the special
conditions of a QCP the effect of additional interactions or
degrees of freedom might be crucial for various physical
properties including the puzzle mentioned above.

Here we report such a new peculiar feature absent to the best of
our knowledge in all other known chain cuprates which should
certainly affect the quantum magnetism. It is specifically related
to the interaction of  the quantum $S=1/2$ spins  in the CuO$_2$
chains with the quantum electrical dipoles derived from tunneling
Li$^+$ ions.  Our $^7$Li nuclear magnetic- (NMR), the Cu$^{2+}$ electron spin
resonance (ESR) and the dielectric constant data show that a
coupling between the active electrical and the magnetic degrees of
freedom occurs already in the paramagnetic regime far above $T_N$.
We argue that this peculiar effect is due to the interaction of
interpenetrating magnetic and electrical sublattices formed
specifically in \Zr.


\begin{figure}[t]

\includegraphics[width=0.85\columnwidth]{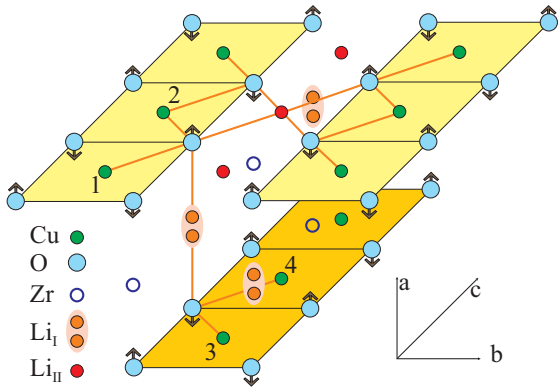}
\caption{(Color online). Fragment of slightly idealized
$\gamma$-\Zr\ structure with the supertransferred
Cu-O-$^7$Li$_{\rm I,II}$ bonds. Arrows show schematically the
oxygen ion displacements in the real structure.} \label{structure}
\end{figure}


The orthorhombic crystal  structure of $\gamma$-\Zr\,($\equiv \rm
Li_2CuZrO_4$)\,~\cite{Dussarrat}  comprises chains formed by
edge-shared CuO$_4$ plaquettes running along the $c$-axis
(Fig.\,\ref{structure}). Owing to an almost 90$^\circ$ Cu-O-Cu
bonding geometry the quantum $S$=1/2 spins of Cu$^{2+}$ ions are
coupled along the chain by the nearest neighbor (NN) ferromagnetic
and the next-NN (NNN) antiferromagnetic exchange interaction
causing spin frustration \cite{Drechsler1}. The CuO$_2$ chains in
\Zr\ form planes similarly to, e.g., LiCu$_2$O$_2$. However, the
interplane distance of 4.7\,\AA \, markedly exceeds that of other
quasi-two dimensional (2D) cuprates. At variance with many other
Li-containing cuprates the Li ions in \Zr\ occupy two different
types of positions: $4b$ (\li) with a 100\,\% occupancy and $8l$
(\lii) with an occupancy of 50\,\% (Fig.\,\ref{structure}). An
unusually large Debye-Waller factor suggests a splitting of the
\lii\ position \cite{Dussarrat}. \vl{This can be modeled by a
double-well potential where the \lii\ ion can hop over the energy
barrier $U_{\mathrm b}$ between two contiguous $8l$ positions.} In
the most pronounced quantum situation \vl{at low temperatures
$k_{\mathrm B}T \ll U_{\mathrm b}$} the \lii\ split site can be
approximated by a two-level system (TLS) described in terms of
tunnelling Li$^+$ ions which is equivalent to pseudospin $s$=1/2
centers with a quantum reorienting electric dipole directed along
the $a$-axis:\,$d_a$=$|e|ds_z$. Here $d$ is the extent of the
\lii\ split position along the $a$-axis. Hence, the \lii\
subsystem can be described as a 3D pseudo-tetragonal
pseudospin-1/2 lattice with \lii\ planes sandwiched between the
adjacent planes of CuO$_2$ chains (Fig.\,\ref{structure}). Thus
\Zr\ may provide a unique model system to study a synergetic
interplay of two interacting low-dimensional quantum subsystems:
the magnetic cuprate chain planes and the \lii-derived electric
dipole planes forming a natural laminar composite structure.

The effective TLS-Hamiltonian for the \lii\  subsystem using the pseudo-spin notation can  be written as follows:

%
\beq \hat H_{Li}=\sum _{i<j}I^{\|}_{ij}{\hat s}_{iz}{\hat s}_{jz} +\hbar\Omega\sum _{i}{\hat s}_{ix}+ \sum _{i<j}I^{\bot}_{ij}{\hat s}_{ix}{\hat
s}_{jx}\,, \label{trising} \eeq

where the first term describes the NN \lii\ -\lii\  interaction, the second one does the \lii\  tunnelling between contiguous $8l$ positions with the
frequency $\Omega$, while the third one does the correlated exchange tunnelling for NN and more distant \lii\ sites. The pseudospin-1/2 Hamiltonian
(\ref{trising}) describes also the well known transversal  field Ising model frequently used in the theory of (anti)ferroelectrics and
quantum glasses. The phase ordering depends essentially on the relation between $\hbar \Omega $ and effective NN coupling energy $I^{\|}_{nn}$,
however, it is an inter-plane frustration and/or the coupling to the spin system that governs the \lii\ dipole ordering. Turning to a magnetoelectric
coupling between the Cu$^{2+}$ planes and  the \lii\ pseudospin-1/2 sublattice one should first note that the \lii\  ion  hopping between two
contiguous $8l$ positions modulates the 2$p$ orbital occupation at nearest oxygen ions  and their displacement. The latter in turn modulates locally
the Cu$^{2+}$ crystal field, the in-plane $t^{\parallel}$ and inter-plane $t^{\perp}$ transfer integrals within a single-band Hubbard-type model
which describes the electron (hole) transfer between the CuO$_4$ plaquettes of the magnetic subsystem. (Note that the magnetic Cu$^{2+}$ sublattice
can also produce an effective electric field which acts on the \lii\  dipoles.) These correlated electronic models can be mapped afterwards on a
Heisenberg model for the Cu spins hence resulting in a modulation of the corresponding exchange integrals. Focussing on one of the \lii\ pseudospin
sites (Fig.~\ref{structure}), both for the Cu$^{2+}$ crystal field $V_{cf}$ and the in-plane Cu$^{2+}$-Cu$^{2+}$ charge transfer we deal with an
anticorrelation effect leading to a local nonequivalence of upper and lower CuO$_2$ chains: $
    \Delta V_{cf}(1)=-\Delta V_{cf}(4)\propto s_z;\,    \Delta V_{cf}(2)=-\Delta V_{cf}(3)\propto s_z $.
The interaction term of the two subsystems reads
\begin{equation}
    \Delta H_{tr}^{\parallel}=\Delta t^{\parallel}(\hat a_1^{\dagger}\hat a_2-\hat a_3^{\dagger}\hat a_4)s_z+h.c.\, ,
\end{equation}
where $\hat a_i^{\dagger}$ and $\hat a_j$ are electronic annihilation and creation operators  and $t^{\parallel}$ denotes the NN-transfer integral in
the chain direction. The modulation of the inter-plane Cu$^{2+}$-Cu$^{2+}$ electron transfer reads as follows
\begin{equation}
\Delta H_{tr}^{\perp}= \Delta t^{\perp}(\sum_{i=1,2;j=3,4}\hat a_i^{\dagger}\hat a_j){\hat s}_x+h.c.\,.
\end{equation}

Mapping the electronic part on a Heisenberg model, we arrive at the famous Kugel-Khomskii model \cite{KK} which describes the interaction of spins
and pseudo-spins usually derived from orbital degrees of freedom. It has been widely used to model mostly on-site interactions of spins and orbitals
and only in the mean-field approximation. To the best our knowledge a case of an {\it inter-site} interaction with a glassy state of one of the
components was not described so far. However, based on our work to be reported below, we argue that it might become relevant for a broad class of
systems with multiple degrees of freedom including orbital physics, too.

To verify the above conjectures we have carried out $^{7}$Li NMR, Cu$^{2+}$ ESR, and dielectric measurements on oriented polycrystals of
Li$_2$CuZrO$_4$. Owing to a small anisotropy of the $g$-factor (see below) it was possible to align powder particles mixed with epoxy resin in a
strong magnetic field. After its hardening a sample with a magnetic "easy" axis parallel to the $a$-axis has been obtained, \vl{as confirmed by the
x-ray diffraction.}

\begin{figure}[t]
\includegraphics[width=0.90\columnwidth]{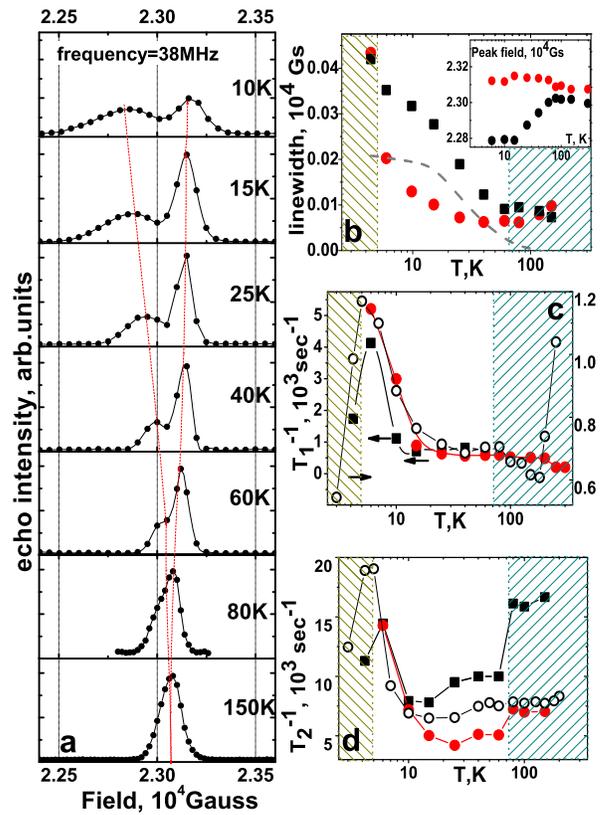}
\caption{(Color online). (a) - Selected $^{7}$Li NMR spectra of the
oriented \Zr \, sample at $\omega_N = 38$\,MHz for ${\bf H}$$\perp$${\bf
a}$; (b)~- $T$-dependence of the NMR linewidth for low- and high-field NMR
lines. Dashed curve denotes an inhomogeneous broadening due to a
glass-like ordering of \lii\ ions. Inset shows the behavior of resonance
fields; (c) and (d) - $T$-dependences of the spin-lattice $T_1^{-1}$ and
spin-spin relaxation rates $T_2^{-1}$, respectively. In (b) - (d)
$\Circle$ denote data for ${\bf H}$$\parallel$${\bf a}$, $\blacksquare$
and $\textcolor{red}{\CIRCLE}$ are data for low- and high-field lines for
${\bf H}$$\perp$${\bf a}$, respectively.} \label{nmr}
\end{figure}

We start with the $^{7}$Li NMR measurements that proved to be very useful before to study both the magnetic ordering in cuprates
\cite{Gippius,Kegler,Tarui08} and the Li ion mobility (see, e.g., Ref.\,\cite{Brinkmann}). The $^7$Li ($I$=3/2) NMR spectra of \Zr \, samples were measured by a Tecmag pulse spectrometer in two orientations: ${\bf H}$$\parallel$${\bf a}$ and ${\bf H}$$\perp$${\bf a}$  by sweeping the magnetic field at a frequency $\omega_N$=38\,MHz. The signal was obtained by integrating the spin-echo envelope.  The longitudinal $T_1^{-1}$ and transversal $T_2^{-1}$ relaxation rates were measured  at the peak of the signal using the stimulated echo sequence and $\pi /2-\pi$ sequence, respectively. The quadrupole splitting, typical for $^7$Li nuclei in cuprates \cite{Gippius,Kegler} and estimated to be of the order of 0.05\,MHz,  is unresolved in the spectrum which shape can be described by a single Gaussian line. In the paramagnetic state at high temperatures ($T\geq 150$\,K $\gg T_N$) the $^7$Li NMR response of \Zr \, (Fig.\,\ref{nmr}), seemingly presents a single line. However, a careful analysis of the lineshape and $T_2^{-1}$ rates shows that we deal with an unresolved superposition of two lines with a full width of 0.01\,T, which may be ascribed to two Li species, i.e. to the immobile \li\ and the mobile \lii\ ions, respectively. For ${\bf H}$$\perp$${\bf a}$ lowering the temperature below $T$$\sim$100\,K yields a separation of the NMR spectrum in two well resolved lines with a strong and unusual $T$-dependent frequency redshift and inhomogeneous broadening of the low field (left) line, (Fig.\,\ref{nmr}), that can be associated with the NMR response of the mobile \lii\ ions. For ${\bf H}$$\parallel$${\bf a}$ the signals from different Li sites merge and only at $T>\vl{T^*} \sim$100\,K one can separate two contributions due to the narrowing of one of them, apparently of \lii. A characteristic temperature $\vl{T^*}\sim$100\,K can be associated with the onset of the quenching of the \lii\  hopping between two equivalent positions \vl{on the NMR timescale}, i.e. with the offset of the motional narrowing. It means that at $T>\vl{T^*}$ the \lii\ pseudospin system reveals most likely a classical high-temperature paraelectric behavior. One should note that the $T$-dependence of the $^7$Li NMR linewidth for immobile \li\ ions (high-field right line) shows up a rather conventional low-$T$ rise due to a critical increase of spin fluctuations in the vicinity of $T_N$. \vl{Since the distances \li--Cu and \lii--Cu are similar, one can assume} roughly the same spin fluctuation contribution to both NMR signals. \vl{Thus} one can single out the additional contribution to the inhomogeneous broadening of the left line (dashed line in Fig.\,\ref{nmr}b) whose $T$-dependence turns out to be typical for systems with mobile Li ions (see, e.g., Ref.\,\cite{Brinkmann}). \vl{The temperature of the ceasing of the motional narrowing of the NMR line is usually correlated with the onset of the critical glassy freezing, which suggests $T^*$ as an upper boundary for the glass ordering temperature $T_{\mathrm g}\lesssim T^*$}.


%
%

The low-$T$ behavior of $T^{-1}_{2}$ (Fig.\,\ref{nmr}d) with characteristic values of $\sim 10^4$\,s$^{-1}$ and a pronounced peak near $T_{N}$ for
both $^7$Li species and both field directions is typical for spin ordering cuprates \cite{Kegler}. The spin-lattice relaxation (SLR) rates $T_1^{-1}$
for both Li species reveal a very similar $T$-behavior with nearly the same constant value of $\sim 750$\,sec$^{-1}$ down to low $T \sim 10$\,K where
both start to increase with a peak at $T \sim 6$\,K for the $^7$\lii\ line and a divergence for the $^7$\li\ line in the case of ${\bf
H}$$\perp$${\bf a}$. A similar behavior is observed for the ${\bf H}$$\parallel$$ {\bf a}$ orientation where contributions from $^7$\li\ and
$^7$\lii\ nuclei are hardly resolved.

A typical mobility induced SLR rate for $^{7}$Li NMR in a wide number of nonmagnetic solids is due to the quadrupole mechanism which is usually small
as compared to a strong SLR due to a magnetic mechanism. In \Zr\ where strong magnetic fluctuations take place in the anisotropic lattice the
decrease of relaxation rates below 100\,K reflects the ceasing of the motion that was averaging local magnetic fields created by fluctuations of Cu
magnetic moments. The anisotropic character of this motion becomes apparent in the different relaxation behavior in different field orientations. One can see in Fig.\,\ref{nmr}c,d that $T_1$ for ${\bf H}\parallel {\bf a}$ increases concomitantly with $T_2$ for ${\bf H}\perp {\bf a}$, which are both caused by magnetic fluctuations perpendicular to the $a$-direction. To summarize this part, all $^7$Li NMR data cleary indicate a freezing of the
\lii\ paraelectric sublattice below $\vl{T^*} \lesssim 100$\,K.

\begin{figure}[t]
\includegraphics[width=0.8\columnwidth]{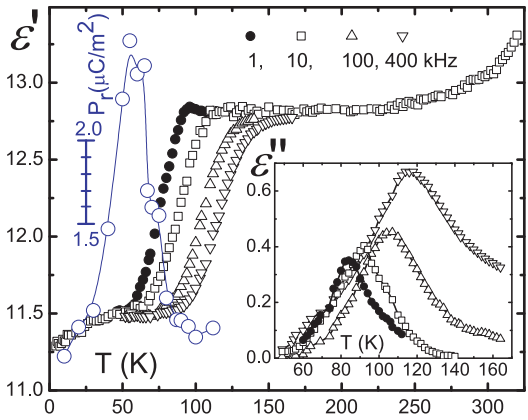}
\caption{(Color online). Temperature dependence of the real ($\varepsilon^{\prime}$) and imaginary ($\varepsilon^{\prime\prime}$) parts of the
dielectric constant for a pressed sample of $\gamma$-Li$_2$CuZrO$_4$ at \vl{different frequencies $f$. $T$-dependence of the remanent polarization
$P_r$ is shown by open circles. Solid line is a guide for the eye.}} \label{epsilon}
\end{figure}

\begin{figure}[t]
\includegraphics[width=0.75\columnwidth, angle=0]{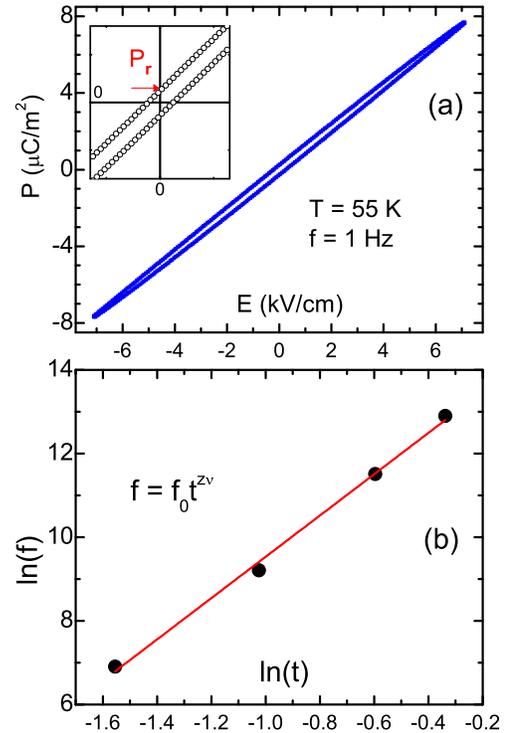}
\caption{(Color online). (a) - Hysteresis loop of the electric field dependence of
the polarization $P(E)$ at $T = 55$\,K.  Inset: enlarged view around the origin with the remanent polarization $P_{\mathrm r}$ indicated by the arrow.
The temperature dependence of $P_{\mathrm r}$ is shown in Fig.\,\ref{epsilon}. \vl{(b) - Full circles - relation between the measurement frequency $f$
and the reduced temperature $t=(T_{\mathrm \epsilon^{\prime\prime}max}/T_{\mathrm g} - 1)$. Solid line - fit to the power law relation $f =
f_{\mathrm 0}t^{z\nu}$with the values of the fluctuation frequency $f_0 = 1.9$\,MHz, the glass ordering temperature $T_{\mathrm g} \simeq 67.4$\,K
and the critical exponent $z\nu = 5.45$ (see the text).}} \label{ar}
\end{figure}

A direct evidence of a glass-like structural ordering of this sublattice can be provided by measurements of the dielectric constant $\varepsilon =
\varepsilon^\prime + i\varepsilon^{\prime\prime}$ that have been performed with pressed pellets of \Zr\ \vl{at a number of frequencies $f$}. The
behavior of the real part $\varepsilon^\prime$ above 300\,K (Fig.\,\ref{epsilon}) evidences most likely a contribution of \lii\  to ionic
conductivity. However, the most \vl{strong} change of $\varepsilon^\prime$ occurs in the $T$-range 50\,-\,150\,K. A significant reduction of
$\varepsilon^\prime$ clearly indicates a 'randomly antiferro' freezing of the paraelectric subsystem in \Zr\ \vl{\cite{Loidl}}. \vl{Remarkably, in
this $T$-range the electric field dependence of the polarization $P(E)$ exhibits a hysteresis (Fig.\,\ref{ar}a) with the values of the remanent
polarization $P_{\mathrm r}$ (Fig.\,\ref{epsilon}) comparable with those of multiferroic cuprates \cite{Naito07,Park07}.}

The frequency dependence of the step in $\varepsilon^\prime(T)$ concomitant with the shift of the peak temperature $T_{\mathrm
\epsilon^{\prime\prime}max}$ of the imaginary part $\varepsilon^{\prime\prime}(T)$  to \vl{lower} $T$ with \vl{decreasing the} frequency $f$ (Fig.\,\ref{epsilon}, inset, and Fig.\,\ref{ar}a) reveals a typical for a glass-like transition \vl{critical dynamic slowing down towards the glass
transition at $T_{\mathrm g}$. The temperature $T_{\mathrm \epsilon^{\prime\prime}max}$ can be associated with the frequency $f$ dependent freezing
temperature below which the longest relaxation time of the system gets larger than the characteristic observation time $1/f$ and the system enters a
nonequilibrium regime \cite{Binder86}. In the three dimensional (3D) case where a finite glass ordering temperature $T_{\mathrm g}$ can be expected,
$f$ and $T_{\mathrm \epsilon^{\prime\prime}max}$ can be connected by the power-law relation $f = f_{\mathrm 0}t^{z\nu}$ \cite{Hohenberg77}. Here
$f_{\mathrm 0}$ is the fluctuation frequency, $t=(T_{\mathrm \epsilon^{\prime\prime}max}/T_{\mathrm g} - 1)$ is the reduced temperature, and $z$ and
$\nu$ are critical exponents. The $f$ vs. $T_{\mathrm \epsilon^{\prime\prime}max}$ dependence can be well fitted by this expression, yielding the
glass ordering temperature $T_{\mathrm g} \simeq 70$\,K and $z\nu \simeq 5.5$, Fig.\,\ref{ar}b. The value of $T_{\mathrm g}$ relates well with the
characteristic NMR 'freezing' temperature $T^*$. Notably, the obtained product of the critical exponents $z\nu \simeq 5.5$ is very close to the
result $z\nu = 6 \pm 1$ of a reported Monte Carlo simulation for the 3D Ising glass \cite{Ogielski85}. This supports the appropriateness of the Ising
(pseudo-spin-1/2) model adopted above for the description of the electric \lii\  subsystem in \Zr.}

\begin{figure}[t]
\includegraphics[width=\columnwidth, angle=0]{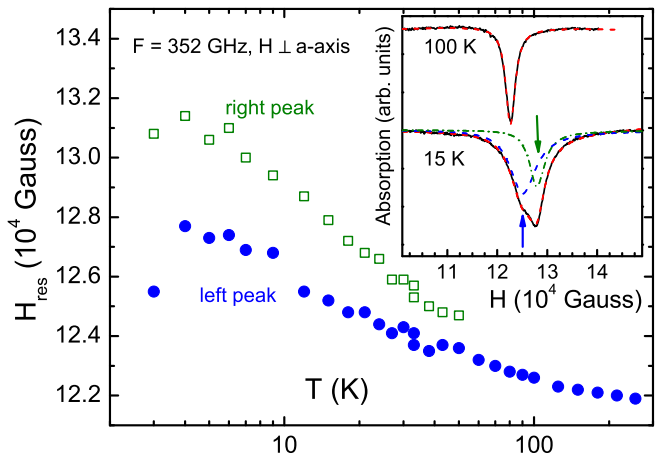}
\caption{(Color online). Inset: HF-ESR spectra at 352\,GHz for ${\bf H}\perp {\bf a}$ at
100\,K and 15\,K, black solid lines. Dash lines (red online) are fits indistinguishable from the experiment. Decomposition of the spectrum at 15\,K
in two lines is shown by dash (blue online) and dash-dot (green online) curves. Main panel: $T$-dependence of the resonance field of the peak(s).}
\label{esr}
\end{figure}

How can the magnetic Cu$^{2+}$ subsystem respond to this effect? First of all one may anticipate  a change of the crystal field and of the
$g$-factors for Cu$^{2+}$ ions which can be detected by the ESR technique. We have carried out high-field ESR (HF-ESR) experiments at sub-THz
frequencies  in a broad temperature range with a home-made spectrometer (for details, see Ref.~\cite{Golze06}). In the high-$T$ regime above \vl{$T_{\mathrm g} \simeq 70$\,K} the ESR spectrum comprising a single Lorentzian absorption line has been found (Fig.\,\ref{esr}, inset). The resonance field corresponds to $g$-factors of $g_{\parallel}$=\,2.19 and $g_{\perp}$=\,2.02 for ${\bf H}$$\parallel$${\bf a}$ and ${\bf H}$$\perp$${\bf a}$, respectively, typical for a Cu$^{2+}$ ion in a distorted square planar ligand coordination\,\cite{Pilbrow}. A remarkable evolution of the ESR spectrum has been observed upon cooling the sample below $T_{\mathrm g}$ where the spectrum begins to split into two lines. The shape of the spectrum can be perfectly fitted with two Lorentzians of comparable intensity, (Fig.\,\ref{esr}), inset. Representative $T$-dependences of the resonance fields at a frequency of 352\,GHz obtained from the fit are shown, Fig.\,\ref{esr}. The two lines in the HF-ESR spectrum whose separation increases with decreasing $T$ can be straightforwardly assigned to the occurrence of two nonequivalent Cu sites with slightly different $g$-factors. Since the $g$-factor is very sensitive to the symmetry and the strength of the ligand electrical field potential $V_{cf}$, the splitting of the spectrum evidences that the mobile \lii\ ions freeze in the lattice on a particular pattern yielding two distinct local crystal field potentials $V_{cf}\pm \Delta V_{cf}$ at the Cu$^{2+}$ sites. The shift of both lines to higher fields indicates the onset of low-frequency spin correlations which develop in a low-D spin system far above $T_N$. Owing to a much shorter timescale of the ESR they are detected at higher $T$ compared to the NMR relaxation data (Fig.~\ref{nmr}c,d).

Finally, we argue that the frozen 'glassy' \lii-derived quantum dipoles may produce a random electrical field potential that might disturb the
oriented 'multiferroic' dipoles induced in some way by the magnetic spiral with the aid of special impurity spins or by the Dzyaloshinski-Moriya
interaction. As a result no multiferroicity will be observed below the magnetic ordering at $T_N$. Furthermore, the absence of any long-range order
of the \lii-dipoles comprising in \Zr\ a {\it regular} (although frustrated) electrical dipole sublattice is noteworthy, because a glassy ground
state was observed so far only for essentially nonstoichiometric random systems \cite{Loidl}. The only feasible source for an effective randomness at
\lii\ sites in our high-quality samples should then be due to short-range spiral-like correlated spins. Note that even a perfectly ordered  but
intrinsically {\it incommensurate} spiral state is expected to introduce a quasi-random potential at the commensurate \lii\ positions. Thus, one
conjectures that owing to a sufficiently strong coupling of spins and pseudospins short-range spin correlations seen by ESR already at
$T$\,$\sim$\,$T_{\mathrm g}$\,$\gg$\,$T_N$ may introduce disorder also in the electrical \lii-dipole sublattice. \vl{An} additional decrease of
$\varepsilon^\prime$ below $\sim$\,30\,K \vl{concomitant with a rapid decrease of $P_{\mathrm r}$ (Fig.\,\ref{epsilon}) may indicate a crossover
from a classical regime dominating the response of the glassy ordered \lii\ sublattice  at higher temperatures to a quantum tunneling regime which
interferes with the development of quantum magnetism in CuO$_2$ chains.} In this context it would be of considerable interest to suppress or to
modulate \vl{the electrical dipole} disorder, e.g., by external electric fields or pressure, and to investigate how the paraspiral state in between
$T_{\mathrm g}$ and  $T_N$ as well as the spiral state below $T_N$ will response. In particular, in the region $T_N < T < T_{\mathrm g}$ the
renormalization of the chain exchange integrals due to the spin-pseudo-spin interaction might take place. As a consequence, it would require
different fits for the magnetic susceptibility $\chi(T)$ \cite{Drechsler1} above and below $T_{\rm g}$, being thus possibly helpful to resolve or to contribute to the solution of the above mentioned $J_1$-puzzle for \Zr. This way, our observation of a pseudo-spin (electric dipole) - spin-1/2
coupling paves a way to new experimental and theoretical studies of \Zr\ to get more insight into the fundamental features of multiferroicity,
magnetic ordering, as well as the nature of the critical point itself.

In conclusion, by measuring  $^{7}$Li NMR, Cu$^{2+}$ ESR and dielectric constants of the quantum spin-1/2 incommensurate chain cuprate \Zr\ we have
observed a new peculiar effect of the interaction between spins-1/2 associated with Cu$^{2+}$ ions and pseudospins-1/2 (Ising-like electrical
dipoles) due to tunneling Li$^+$ ions. From the  NMR  and dielectric measurements we conclude that at $T_{\mathrm g} \vl{\simeq 70}$\,K a glass-like
structural ordering of the electrical sublattice \vl{takes place}. Its remarkable impact on the spin system is evidenced by the development of spin
site nonequivalency at $T < T_{\mathrm g}$ revealed by ESR. These results put forward \Zr\ \vl{as new type of the intrinsic single-phase
magnetoelectric composites where quantum magnetism of CuO$_2$ chains meet the electrically active regular sublattice of Li$^+$ ions. As such it
emerges} as a unique model system to study the influence of additional interactions and degrees of freedom on frustrated magnetism near a critical
point where the magnetic spin subsystem is especially soft. One of the feasible consequences of such influence could be, e.g., the suppression of
multiferroicity in the title compound.

Support of the DFG (grants 436 RUS 113/936/0-1, \vl{DR269/3-1 and KL1824/3-1}) and the RFBR (grants 08-02-91952-NNIO-a, 07-02-01184-a, 06-02-17242,
07-02-96047, and 08-02-00633), is gratefully acknowledged. YA acknowledges support of the EU Programme Alban.  We thank A.A. Zvyagin and N.M. Plakida for discussions.

\end{document}